\documentclass[aps,prl,floatfix,twocolumn,showpacs,amsmath,amssymb,superscriptaddress]{revtex4}

\usepackage{bm,enumerate,amssymb}
\usepackage{comment,color}
\usepackage{graphicx}
\usepackage{subfigure}
\usepackage{hyperref}

\newcommand{\beq}{\begin{equation}}
\newcommand{\eeq}{\end{equation}}
\newcommand{\beqn}{\begin{eqnarray}}
\newcommand{\eeqn}{\end{eqnarray}}

\def\kappaETfull{{$\kappa$-(BEDT-TTF)$_{2}$Cu$_{2}$(CN)$_{3}$}}
\def\dmitfull{{EtMe$_{3}$Sb[Pd(dmit)$_{2}$]$_{2}$}}

\def\kappaET{{$\kappa$-BEDT}}
\def\dmit{{DMIT}}

\newcommand{\rvm}[1]{#1}

\date{July 22, 2013}

\begin{document}

\title{\rvm{Theory of a competitive spin liquid state for weak Mott insulators\\ on the triangular lattice}}

\author{Ryan V.\ Mishmash}
\affiliation{Department of Physics, University of California,
Santa Barbara, California 93106, USA}

\author{James R.\ Garrison}
\affiliation{Department of Physics, University of California,
Santa Barbara, California 93106, USA}

\author{Samuel Bieri}
\affiliation{Department of Physics, Massachusetts Institute of
Technology, Cambridge, Massachusetts 02139, USA}

\author{Cenke Xu}
\affiliation{Department of Physics, University of California,
Santa Barbara, California 93106, USA}

\begin{abstract}

We propose a novel quantum spin liquid state that can explain many
of the intriguing experimental properties of the low-temperature
phase of the organic spin liquid candidate materials
\rvm{\kappaETfull~and~\dmitfull.} This state of paired fermionic
spinons preserves all symmetries of the system, and it has a
gapless excitation spectrum with quadratic bands that touch at
momentum $\vec{k}=0$. This quadratic band touching is protected by
symmetries. Using variational Monte Carlo techniques, we show that
this state has highly competitive energy in the triangular lattice
Heisenberg model supplemented with a realistically large
ring-exchange term.

\end{abstract}

\pacs{71.27.+a, 75.10.Jm, 75.10.Kt, 75.30.Kz}

\maketitle

Quantum spin liquids are exotic phases of quantum spin systems
which break no global symmetries even when thermal fluctuations
are completely suppressed at zero
temperature~\cite{Balents10_Nature_464_199,Lee08_Science_321_5894}.
In the last decade, candidates of gapless spin liquid phases have
been discovered in various experimental systems, \rvm{including
\kappaETfull~\cite{kappathermo1,Shimuzi03_PRL_91_107001,Kurosaki05_PRL_95_177001},
\dmitfull~\cite{131dmit1,131dmit2,131dmit3},
$\mathrm{Ba_3CuSb_2O_9}$~\cite{haidongcu},
$\mathrm{Ba_3NiSb_2O_9}$~\cite{haidongni}, and
$\mathrm{ZnCu_3(OH)_6Cl_2}$~\cite{kagome,kagome2,PhysRevLett.109.037208}.}
In all these materials, no evidence of magnetic order was found at
temperatures much lower than the spin interaction energy scale of
the system. In the current work, we will focus on the organic spin
liquid materials \kappaETfull~(\kappaET) and \dmitfull~(\dmit).
These materials are
quasi-two-dimensional 
Mott insulators which are close to a Mott metal-insulator
transition~\cite{Kurosaki05_PRL_95_177001}, and thus exhibit
substantial local charge fluctuations. An effective spin model
that may well describe the magnetic properties of these ``weak''
Mott insulators involves supplementing the usual (possibly
extended) Heisenberg model with a four-site ring-exchange term
\cite{Motrunich05_PRB_72_045105,Misguich99_PhysRevB.60.1064}.
Here, we consider the following Hamiltonian:
\begin{equation}
H = J_1\sum_{\langle i,j\rangle} 2\vec{S}_i \cdot \vec{S}_j + J_2
\sum_{\langle\langle i,j\rangle\rangle} 2\vec{S}_i \cdot \vec{S}_j
+ K\sum_{\langle i,j,k,l\rangle} (P_{i j k l} + \text{H.c.}),
\label{eq:modelJK}
\end{equation}
where the sums $\langle i,j\rangle$ and $\langle \langle i,j
\rangle \rangle$ go over all first- and second-neighbor links of
the triangular lattice, respectively, while $\langle
i,j,k,l\rangle$ goes over all elementary four-site rhombi;
$P_{ijkl}$ rotates the spin configurations around a given rhombus.
In what follows, we set $J_1=1$ as the unit of energy.

The two different organic spin liquids \kappaET~and \dmit~share
two universal properties:

{\it 1.} At low temperatures, despite the fact that the system is
still a Mott insulator for charge transport, the specific heat
scales linearly with temperature: $C_v = \gamma T$. Furthermore,
$\gamma$ is essentially independent of a moderate external
magnetic field~\cite{kappathermo1}.

{\it 2.} The spin susceptibility shows no magnetic phase
transition at any finite temperature, and it saturates to a finite
constant $\chi$ at zero
temperature~\cite{Shimuzi03_PRL_91_107001}.

These two phenomena are completely inconsistent with any
semiclassical magnetic state and are strongly suggestive of the
existence of a highly nontrivial quantum disordered phase.
\rvm{They also demonstrate} the presence of a large density of
charge-neutral excitations at low temperature.  To date, four main
theoretical scenarios have been proposed to describe these
experimental facts:

{\it 1.} In the U(1) spinon Fermi surface
state~\cite{Motrunich05_PRB_72_045105,Lee05_PRL_95_036403}, a
fermionic spinon $f_{j\alpha}$ is introduced by decomposing the
physical spin operator as $\vec{S}_j = \frac{1}{2} \sum_{\alpha,
\beta = \uparrow, \downarrow} f^\dagger_{j\alpha}
\vec{\sigma}_{\alpha\beta} f_{j\beta}$ and taking the spinons to
fill an ordinary Fermi sea at the mean-field level. This gives
rise to a finite density of states, consistent with the
experimental results mentioned above.  Furthermore, it has been
demonstrated that for strong enough ring exchange $K$, the spinon
Fermi sea state has very competitive variational energy in the
microscopic spin model
(\ref{eq:modelJK})~\cite{Motrunich05_PRB_72_045105}. However, once
we go beyond the mean-field level, the U(1) gauge fluctuation will
acquire singular overdamped dynamics $|\omega| \sim k^3$ due to
its coupling with the Fermi surface~\cite{polchinski}. This
singular dynamics generates an even larger density of states at
low temperature, which leads to a singular specific heat $C_v \sim
T^{2/3}$. This specific heat behavior is not observed
experimentally.

{\it 2.} The most natural way to suppress the U(1) gauge
fluctuation is to condense Cooper pairs of spinons and thus break
the U(1) gauge fluctuation down to a fully gapped $Z_2$ gauge
fluctuation.  This possibility has been explored numerically in
Ref.~\cite{Grover10_PRB_81_245121}, where the authors concluded
that the particular pairing pattern that is energetically favored
by Eq.~(\ref{eq:modelJK}) has nodal $d_{x^2-y^2}$-wave structure.
However, this nodal $d$-wave pairing not only suppresses the gauge
fluctuation, it also significantly suppresses the fermion density
of states, and the system will no longer have finite $\gamma$ and
$\chi$ at low temperature, unless sufficient disorder is turned
on.



{\it 3.} Another very different approach was taken in
Ref.~\cite{qixu}, where the authors proposed that \kappaET~is a
$Z_2$ spin liquid which is very close to the condensation quantum
critical point of bosonic spinons.  This quantum critical behavior
is consistent with the NMR relaxation rate observed
experimentally~\cite{kappa2}. In particular, the small energy gap
seen in thermal conductivity data~\cite{kappathermo2} was
identified with the gap of the topological defect of the $Z_2$
spin liquid~\cite{qixu}.  However, no parent spin Hamiltonian has
been found for this state so far. Thus, it is unknown whether this
quantum critical spin liquid can be realized in any experimentally
relevant lattice model.

{\it 4.} A novel Majorana slave fermion formalism was introduced
in Ref.~\cite{rudro}, where the authors proposed that the ground
state of the organic spin liquids has a Majorana Fermi surface.
But, just like the previous theory, so far it is unclear in which
lattice model this spin liquid can be realized.

\begin{figure}[t]
\includegraphics[width=0.9\columnwidth]{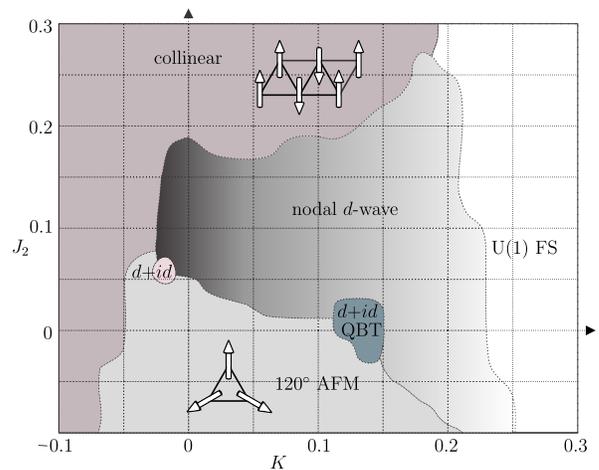}
\caption{Variational phase diagram of the spin Hamiltonian,
Eq.~(\ref{eq:modelJK}).\label{phasediagram} We propose that the
$d+id$ QBT spin liquid phase is a very strong candidate for the
ground state of \kappaET~and \dmit~in the parameter range
$J_2\simeq 0$ and $K\simeq 0.13$.}
\end{figure}

In this paper, we propose an entirely new spin liquid. In
Ref.~\cite{kivelson}, possible $Z_2$ spin liquids with an extended
spinon Fermi surface were summarized. However, the spin liquid
state proposed in the present paper is beyond the ones discussed
in Ref.~\cite{kivelson}. Our novel state has no spinon Fermi
surface, but has a quadratic band touching (QBT) of fermionic
spinons that is protected by the symmetry of the model: $\omega
\sim \pm k^2$. In two dimensions, a quadratic band touching leads
to a finite constant density of states, which automatically gives
finite $\gamma$ and $\chi$ at zero temperature. Besides being
consistent with the major experimental facts of the organic spin
liquid compounds, this state has the following advantages:

{\it 1.} As we will show below, this state is a very competitive
variational ground state for model~(\ref{eq:modelJK}) in the
physically relevant regime $0.1\lesssim K\lesssim0.15$ and
$J_2\simeq0$ (see Fig.~\ref{phasediagram}).

{\it 2.} The gauge fluctuation for this state is fully gapped, and
hence plays no role at low energy. Most field-theoretic
calculations based on this state are thus well approximated at the
mean-field level, and so, in contrast to the spinon Fermi surface
state \cite{Lee09_PRB_80_165102}, they are well controlled.

{\it 3.} Finite $\gamma$ and $\chi$ are \emph{generic} properties
of our QBT spin liquid. In
 contrast to the spinon Fermi surface state,
these properties are \rvm{both robust in the presence of} gauge
fluctuations, and unlike the nodal $d$-wave state, they do not
rely on disorder.

{\it 4.} A very small energy gap, much smaller than the Heisenberg
exchange $J_1$, was observed by thermal conductivity measurements
in \kappaET~\cite{kappathermo2}. This small gap can be very
elegantly explained by our QBT spin liquid without fine-tuning: an
allowed short-range spinon interaction on top of our mean-field
state may be marginally relevant, and thus naturally open up an
exponentially small gap.

{\it 5.} Since the gauge field fluctuation is fully gapped in our
spin liquid, it does not respond to an external magnetic field.
Thus our state has no obvious thermal Hall effect, which is
consistent with experiments~\cite{131dmitthermo2}.

Let us first describe the QBT spin liquid state. We take the
standard slave fermion (spinon) representation of spin-1/2
operators: $\vec{S}_j = \frac{1}{2} \sum_{\alpha, \beta =
\uparrow, \downarrow} f^\dagger_{j\alpha}
\vec{\sigma}_{\alpha\beta} f_{j\beta}$. The physical spin-1/2
Hilbert space is then recovered by imposing the on-site constraint
$\sum_\alpha f^\dagger_{j\alpha}f_{j\alpha} = 1$, which introduces
an SU(2) gauge symmetry to the low-energy dynamics of the
spinons~\cite{wen2002}. However, this SU(2) gauge symmetry will
generally be broken by the mean-field dynamics, which can be
described by a quadratic Hamiltonian of the form \beqn
\label{eq:Hmf} H_\mathrm{MF} = -\sum_{i,j}\left[t_{ij} f_{i
\sigma}^\dagger f_{j \sigma} + \left(\Delta_{ij} f^\dagger_{i
\uparrow} f^\dagger_{j \downarrow} + \text{H.c.}\right) \right]\,.
\eeqn

The QBT spin liquid at the focus of this paper corresponds to a
mean-field ansatz for the spinons with $d+id$ pairing and
vanishing hopping: \beqn t_{ij} = 0, \ \ \ \Delta_{j,j+\hat{e}} =
\Delta (e_x+i e_y)^2\,. \label{ansatz} \eeqn Here, $\hat{e}$ is a
first-neighbor unit vector of the triangular lattice.
This mean-field ansatz breaks the SU(2) gauge symmetry down to
$Z_2$: $f_\alpha \mapsto - f_{\alpha}$.  Thus, gauge fluctuations
can be ignored in the low-energy dynamics of the system.

It is convenient to introduce a complex spinor $\psi$ defined as
$(\psi_1, \ \psi_2) = (f_{\uparrow}, \ f^\dagger_{\downarrow})$.
Expanded at the $\Gamma$-point $\vec{k} = 0$, the low-energy
Hamiltonian for the mean-field ansatz mentioned above reads \beqn
H_0 = \psi^\dagger \{ - \tau^x (\partial_x^2 - \partial_y^2) + 2
\tau^y
\partial_x
\partial_y \} \psi\,. \label{mfqbt}\eeqn
This mean-field Hamiltonian has a quadratic band touching at
$\vec{k} = 0$, which leads to a finite density of states in two
dimensions. We propose that this finite density of states is
responsible for finite $\gamma$ and $\chi$ observed experimentally
in \kappaET~and \dmit. A similar QBT spin liquid state for the
spin-1 material $\mathrm{Ba_3NiSb_2O_9}$~\cite{haidongni} was
proposed in Ref.~\cite{Cenke12_PRL_108_087204r}.

In addition to the quadratic band touching at $\vec{k} = 0$, there
are also Dirac fermions at the corners of the Brillouin zone:
$\vec{Q}_{A,B} = \pm (4\pi/3, 0)$. Complex Dirac fermion fields
$\psi_{A,B}$ at momenta $\vec{Q}_{A,B}$ can be defined as $\psi =
\psi_A \exp(i \vec{Q}_A \cdot \vec{r}) + \psi_B \exp(i \vec{Q}_B
\cdot \vec{r})$. The low-energy Hamiltonian for $\psi_{A,B}$ reads
\beqn H_{\pm (4\pi/3, 0)} = \sum_{a = A, B} \psi^\dagger_{a}(- i
\tau^x
\partial_x - i \tau^y \partial_y )\psi_a\,. \label{mfdirac} \eeqn
At low temperature, the contribution of these Dirac fermions to
$\gamma$ and $\chi$ is much smaller than the one resulting from
the quadratic band touching at the $\Gamma$-point.

The spinon carries a projective representation of the physical
symmetry group. In the Supplemental Material, we demonstrate that
the mean-field QBT ansatz discussed above preserves all the
symmetries of the model (including the spin symmetry, triangular
lattice symmetry and time-reversal symmetry).  As long as these
symmetries are preserved, no relevant fermion bilinear terms can
be added to Eqs.~(\ref{mfqbt}) and (\ref{mfdirac}), and the
low-energy dynamics is stable.

Let us now go beyond the mean field. As mentioned above, the
mean-field ansatz breaks the gauge symmetry down to $Z_2$, and the
gauge fluctuations are thus quite harmless. But, besides the gauge
fluctuation, local short-range four-fermion interactions exist at
both the Dirac points $\vec{Q}_{A,B}$ and the QBT $\Gamma$-point.
At the $\Gamma$-point, only the following four-fermion interaction
needs to be considered: \beqn H_4 = - g f^\dagger_\uparrow
f_\uparrow f^\dagger_\downarrow f_\downarrow \sim g \psi^\dagger_1
\psi_1 \psi^\dagger_2 \psi_2\, .\eeqn The RG flow of this term is
very simple: depending on the sign of $g$, $H_4$ can be either
marginally relevant or irrelevant~\cite{sunyao}. When it is
relevant ($g > 0$), the system spontaneously breaks time-reversal
symmetry (it becomes a chiral spin liquid) and opens up an
exponentially small gap at the $\Gamma$-point: $m
f^\dagger_{\alpha} f_{\alpha} = m \psi^\dagger \tau^z \psi$. We
identify this fluctuation generated gap with the small gap
observed by thermal conductivity in \kappaET~\cite{kappathermo2}.

In \dmit, on the other hand, thermal conductivity measurements
indicate that the system is gapless at the lowest temperature
\cite{131dmitthermo2}. Thus, we conjecture that \dmit~corresponds
to the case with a marginally irrelevant $H_4$ ($g < 0$), while
\kappaET~ corresponds to $g > 0$.  The thermal conductivity
behavior with negative $g$ will be studied in detail in the
future~\cite{xumoonfuture}, taking into account both interaction
and disorder effects.

Inspired by previous works \cite{Motrunich05_PRB_72_045105,
Grover10_PRB_81_245121, Sheng09_PRB_79_205112,
Block11_PRL_106_157202, GrosRVB}, we now revisit the variational
phase diagram of model~(\ref{eq:modelJK}) using a wide range of
correlated spin wave functions. The quadratic Hamiltonian,
Eq.~(\ref{eq:Hmf}), allows straightforward construction of spin
liquid wave functions by Gutzwiller projecting its ground state
$|\Psi_0\rangle$. That is, we use as variational states
$|\Psi(\{t_{ij}\}, \{\Delta_{ij}\})\rangle = \mathcal{P}_G
\mathcal{P}_N|\Psi_0\rangle$, where $\mathcal{P}_N$ is a projector
to a state with $N$ spinons, and $N$ is the number of lattice
sites ($N_\uparrow=N_\downarrow=N/2$). $\mathcal{P}_G = \prod_j[1
- n_{j\downarrow}n_{j\uparrow}]$ is the Gutzwiller projector which
removes unphysical states containing doubly-occupied sites. We fix
the spinon chemical potential $\mu=t_{jj}$ such that
$|\Psi_0\rangle$ is half filled on average before projection, but
other parameters in (\ref{eq:Hmf}) are used as variational
parameters. The evaluation of expectation values in such fermionic
wave functions can be done efficiently and with high accuracy
using variational Monte Carlo techniques
\cite{GrosRVB,Ceperley77_PRB_16_3081, Bieri12_PRB_86_224409}. For
competing long-range ordered states, we use Jastrow-type wave
functions as pioneered by Huse and Elser \cite{Huse88_PRL_60_2531}
(see the Supplemental Material for more details on all states we
studied).

We first consider the case with $J_2 = 0$ in
Eq.~(\ref{eq:modelJK}).
Since the seminal work of Motrunich
\cite{Motrunich05_PRB_72_045105} it has been known that the U(1)
projected Fermi sea state (or ``spin Bose metal''
\cite{Sheng09_PRB_79_205112}) with isotropic nearest-neighbor
$t_{ij} = t$ and $\Delta_{ij}=0$ has remarkably good variational
energy and is clearly the best fermionic trial state for
relatively large ring exchange $K\gtrsim0.3$. This state is also
consistent with recent large-scale DMRG calculations on the
four-leg ladder \cite{Block11_PRL_106_157202}. On the other hand,
exact diagonalization studies \cite{LiMing00_PRB_62_6372} indicate
that the $120^\circ$ antiferromagnetic (AFM) order, which is
believed to characterize the ground state of the Heisenberg model
at $K=0$ \cite{Huse88_PRL_60_2531, Capriotti99_PRL_82_3899,
White07_PRL_99_127004}, is destroyed at much smaller ring exchange
$K\gtrsim 0.1$. Therefore, an intermediate spin liquid phase in
the parameter regime $0.1\lesssim K \lesssim 0.3$ may well be
present in the model, and is likely to be relevant for the organic
compounds.

The most natural candidate states are $Z_2$ spin liquids with
finite spinon pairing $\Delta_{ij}\neq 0$ in~(\ref{eq:Hmf}).
Indeed, it has been known since the work of Motrunich that in the
intermediate parameter regime of Eq.~(\ref{eq:modelJK}) such
projected Bardeen-Cooper-Schrieffer states do have significantly
lower energy than the $120^\circ$ AFM and U(1) Fermi sea states.
However, the nature of the spinon pairing pattern in this putative
$Z_2$ spin liquid was still up for debate. In this paper, we
perform accurate large-scale simulations up to $30\times30$
lattice sites to check all singlet ($\Delta_{ij} = \Delta_{ji}$)
and triplet ($\Delta_{ij} = -\Delta_{ji}$) pairing instabilities
($s$, $p$, $p+ip$, $d$, $d+id$, and $f$-wave) of the U(1) Fermi
sea state in model (\ref{eq:modelJK}). We find the remarkable
result that for $0.1\lesssim K \lesssim 0.15$ our QBT $d+id$
state, as discussed above, is highly competitive, and perhaps has
the lowest energy of \emph{any} projected fermionic trial state,
including the nodal $d$-wave state of
Ref.~\cite{Grover10_PRB_81_245121}.

\begin{figure}[b]
\begin{center}
\subfigure{\includegraphics[height=0.55\columnwidth]{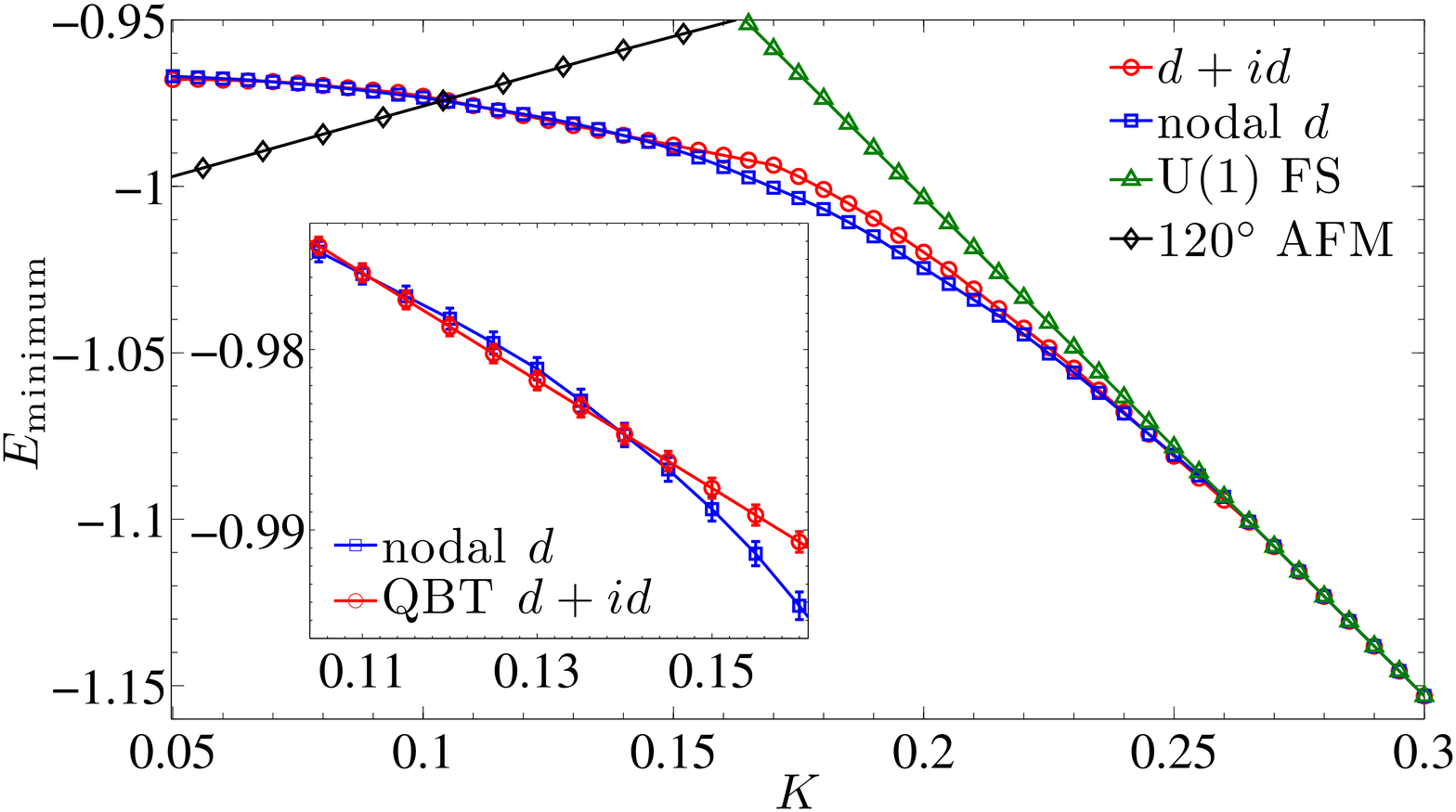}}
\end{center}
\vspace{-0.6cm}
\begin{center}
\hspace{0.23cm}\subfigure{\includegraphics[height=0.54\columnwidth]{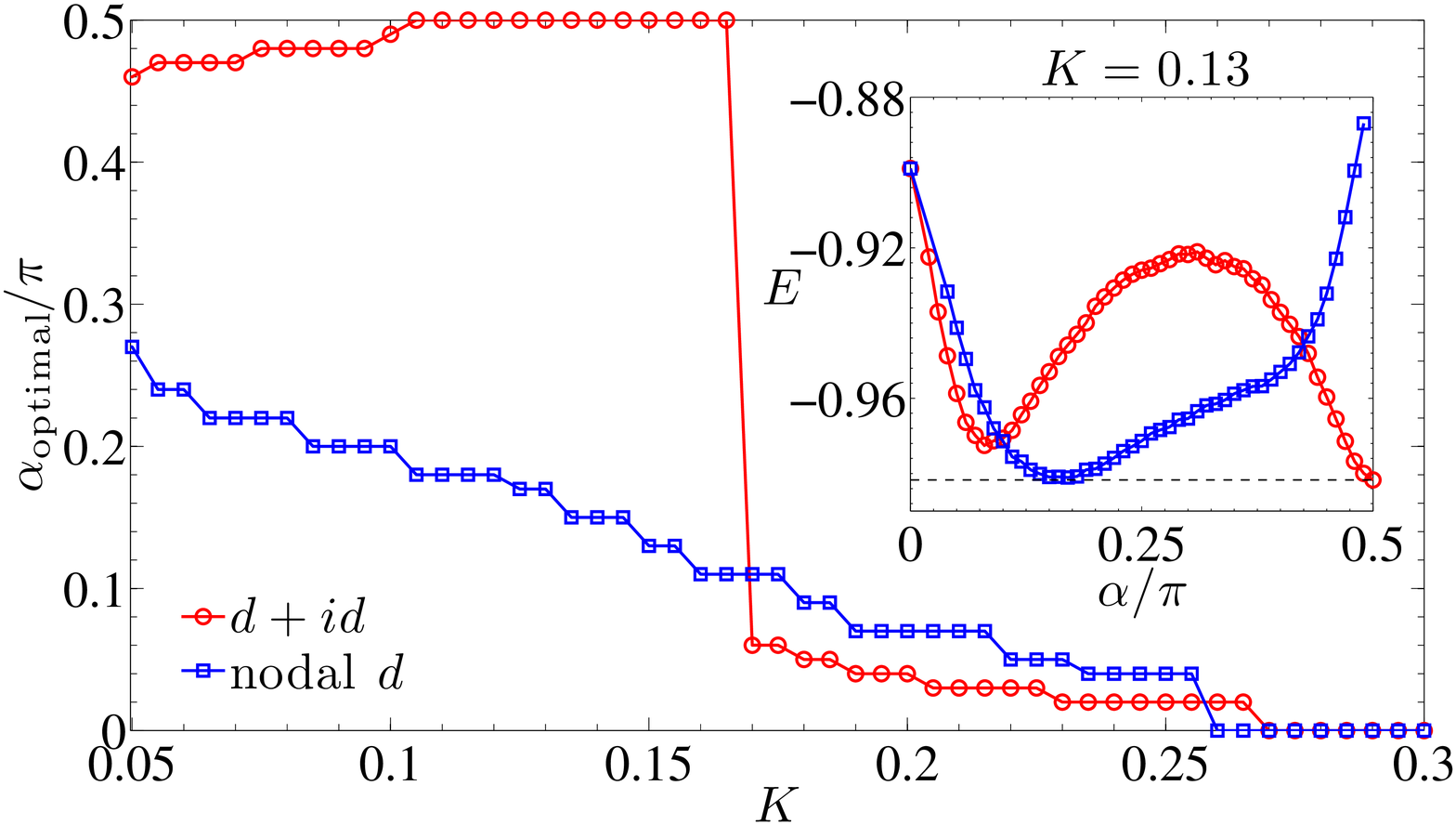}}
\end{center}
\vspace{-0.4cm} \caption{Upper panel:  Variational energies per
site for the Hamiltonian, Eq.~(\ref{eq:modelJK}), at $J_2=0$ as a
function of $K$ for the most competitive trial states in our
study; in the inset, we show a zoom of the region of the phase
diagram where the QBT $d+id$ state is most competitive. Lower
panel:  The optimal variational parameter $\alpha
=\tan^{-1}(\Delta/t)$ is plotted for the $d+id$ and nodal $d$-wave
states; in the inset, we show the variational energies for all
$\alpha$ at the point $K=0.13$ where the dashed line indicates the
energy of the QBT state.} \label{fig:summary_J2_0}
\end{figure}

The results of our variational study at $J_2=0$ are summarized in
Fig.~\ref{fig:summary_J2_0}. Consistent with
Refs.~\cite{Motrunich05_PRB_72_045105, Grover10_PRB_81_245121}, we
find that the unpaired U(1) Fermi sea (FS) state and states with
nodal $d$-wave and $d+id$ pairing symmetries are the most
competitive spin liquid wave functions for this model. The gap
functions for the $d+id$ and nodal $d$-wave states are given by
$\Delta_{j,j+\hat{e}}^{(d+id)}=\Delta\left(e_x+i e_y\right)^2$ and
$\Delta_{j,j+\hat{e}}^{(\mathrm{nodal}~d)}=\Delta\left(e_x^2-e_y^2\right)$,
respectively, where $\hat{e}$ is a unit vector connecting nearest
neighbors on the triangular lattice.
Each of these ans\"atze thus has one variational parameter
$\Delta/t$ which we parameterize by $\alpha=\tan^{-1}(\Delta/t)$.
In the top panel of Fig.~\ref{fig:summary_J2_0}, we show the
optimal energies per site, $E_\mathrm{minimum}$\,, versus ring
exchange $K$ for the $d+id$, nodal $d$-wave, U(1) FS, and
$120^\circ$ AFM states.
In the bottom panel, we show the corresponding optimal $\alpha$
for both the $d+id$ and nodal $d$-wave states. We see that the
$120^\circ$ AFM state wins for $K\lesssim0.1$; however,
immediately upon exiting the $120^\circ$ phase for $0.1\lesssim
K\lesssim 0.15$, the $d+id$ and nodal $d$-wave states are
extremely close in energy and are basically degenerate within
statistical error. Remarkably, as seen in the bottom panel of
Fig.~\ref{fig:summary_J2_0}, the optimal $d+id$ state in the
entire range $0.1\lesssim K\lesssim 0.17$ is in fact our exotic
QBT state of interest, that is $\Delta/t\rightarrow\infty,
\alpha=\pi/2$. For still larger $K$, $0.15\lesssim K \lesssim
0.25$, the optimal ansatz is the nodal $d$-wave state, a result
which is consistent with Ref.~\cite{Grover10_PRB_81_245121}.
Finally, for $K\gtrsim0.25$, the optimal pairing amplitude
$\Delta\rightarrow 0$ for all spin liquid states, thus describing
a crossover to the U(1) Fermi sea state of
Refs.~\cite{Motrunich05_PRB_72_045105, Sheng09_PRB_79_205112,
Block11_PRL_106_157202}.

In the inset of the bottom panel of Fig.~\ref{fig:summary_J2_0},
we plot the variational energies per site versus $\alpha$ at the
point $K=0.13$ in the spin liquid phase. Interestingly, there are
two local minima for the $d+id$ ansatz: the first minimum at small
$\Delta\lesssim t$ is smoothly connected to the U(1) Fermi sea
state at $\Delta=0$, while the second minimum at
$\Delta/t\rightarrow\infty$ is the qualitatively new QBT state at
the focus of our work. For $0.1\lesssim K \lesssim 0.17$, the
latter is lower in energy than the former, but is almost
degenerate with the optimal nodal $d$-wave state which always has
small $\Delta\lesssim t$. Indeed, the two local minima in the
$d+id$ ansatz are already present in the pure Heisenberg model
($K=J_2=0$), with a \emph{large}-$\Delta$ state ($\alpha=0.44$,
$\Delta/t=5.2$) having minimum energy. Furthermore, the QBT state
at $\alpha=\pi/2$ has surprisingly low ring-term expectation
value, and this conspires with the good Heisenberg energy of the
generic large-$\Delta$ $d+id$ state to make the QBT state the
optimal fermionic spin liquid ansatz in the parameter window
$0.1\lesssim K\lesssim 0.15$. (For more details, see the
Supplemental Material.)

The authors of Ref.~\cite{Grover10_PRB_81_245121} concluded that
the nodal $d$-wave state is clearly the best variational ground
state for intermediate $0.1\lesssim K\lesssim 0.15$. We believe
that there are two reasons for this discrepancy with our result.
First, Ref.~\cite{Grover10_PRB_81_245121} considered only a
restricted range of small $\Delta/t$ for the $d+id$ ansatz which
excluded the QBT state altogether. Second, our extensive
finite-size analysis shows that quite large lattice clusters
($\gtrsim 18\times18$ sites) are necessary to get well converged
expectation values for the nodal $d$-wave state. Our calculations
find poorly converged expectation values and strong dependencies
on the spinon boundary conditions for a nodal $d$-wave state on
the small $10\times11$ cluster that was used in
\cite{Grover10_PRB_81_245121}.

Finally, we discuss the effect of a second-neighbor interaction
$J_2$. In Fig.~\ref{phasediagram}, we present a variational phase
diagram in the $K$-$J_2$ plane. A ferromagnetic interaction
($J_2<0$) quickly favors the $120^\circ$ AFM state over the QBT
$d+id$ state and destroys the spin liquid phase.
On the other hand, antiferromagnetic $J_2>0$ strongly frustrates
the 120$^\circ$ AFM state and favors a nodal $d$-wave spin liquid.
Negative ring-exchange or larger values of $J_2\gtrsim 0.17$ lead
to a collinear phase. In Fig.~\ref{phasediagram}, around \rvm{$J_2
\simeq 0.05$} and $K \simeq -0.02$, a small fully gapped $d+id$
phase with finite $\Delta/t$ emerges. This is a chiral spin liquid
with nontrivial topological order~\cite{chiral1,chiral2}. Our
preliminary results show that this phase will expand significantly
once an antiferromagnetic third-neighbor Heisenberg coupling $J_3$
is added to Eq.~(\ref{eq:modelJK}). More details on this phase
will be elaborated in future work.

The ability of the nodal $d$-wave state to beat the collinear
state for $K\simeq0$ may suggest (see
Refs.~\cite{Dagotto90_PRB_42_4800, Lecheminant95_PRB_52_6647})
that we are overestimating the extent of the nodal $d$-wave state
in our phase diagram (see also the Supplemental Material). Of
course, a variational study can never claim to have the final say
on the phase diagram of a given microscopic model, and
quantitative locations of phase boundaries should not be taken too
seriously. What is very robust, however, is the statement that our
QBT $d+id$ state has both extremely competitive energetics in a
realistic parameter regime and highly appealing phenomenology for
the organic spin liquid compounds.

The authors would like to thank T.~Grover, P.~A.~Lee, T.~Senthil,
B.~K.~Clark, and M.~P.~A.~Fisher for helpful discussions. We are
especially grateful to O.~I.~Motrunich for enlightening
discussions and for sharing of his early variational Monte Carlo
data. We also thank Fa~Wang for discussions at the beginning of
this project. R.V.M. and J.R.G. are supported by NSF grant
DMR-1101912. S.B. is supported by NSF grant DMR-1104498. C.X. is
supported by NSF grant DMR-1151208 and the Packard Fellowship.
This work was made possible by the computing facilities of the
Center for Scientific Computing from CNSI, MRL (NSF MRSEC award
DMR-1121053), and NSF grant CNS-0960316; and by the computing
cluster of the MIT Physics Department.

\bibliographystyle{prsty}
\bibliography{QBTdid_biblio}

\appendix{}

\section{Supplementary Materials}

\section{A: Projective symmetry group analysis of the QBT $d+id$ $Z_2$ spin liquid}

We take the standard slave fermion representation of the spin-1/2
operator at site $j$: \beqn \hat{S}^\mu_j = \frac{1}{2}
\sum_{\alpha, \beta = \uparrow, \downarrow} f^\dagger_{j\alpha}
\sigma^\mu_{\alpha\beta} f^{\vphantom\dagger}_{j\beta}. \eeqn The
gauge symmetry of this representation is SU(2), denoted as
SU(2)$_g$. In order to make both spin SU(2) symmetry [SU(2)$_s$]
and gauge SU(2) symmetry manifest, it is convenient to define the
following Majorana fermion $\eta$: \beqn f_{j\alpha} =
\frac{1}{2}(\eta_{j,\alpha, 1} + i\eta_{j,\alpha, 2}). \eeqn

On every site, $\eta_j$ has in total two two-component spaces,
making the maximal possible transformation on $\eta_j$ SO(4).
Within this SO(4), the SU(2)$_s$ transformations and SU(2)$_g$
transformations are generated by the following operators: \beqn
\mathrm{SU(2)}_s &:& (\sigma^x\lambda^y, \ \ \sigma^y, \ \
\sigma^z \lambda^y ), \cr\cr \mathrm{SU(2)}_g &:&
(\sigma^y\lambda^z, \ \ \sigma^y\lambda^x, \ \ \lambda^y ), \eeqn
where the Pauli matrices $\lambda^a$ operate on the two-component
space $(\mathrm{Re}[f], \mathrm{Im}[f])$. SU(2)$_s$ and SU(2)$_g$
commute with each other.

In terms of slave fermions $f_{\alpha}$, the Heisenberg model
reads \beqn \sum_{i, j, \mu} J_{ij} \hat{S}^\mu_{i}\hat{S}^\mu_j
&\sim& \sum_{i, j, \mu} J_{ij} f^\dagger_{i\alpha}
\sigma^\mu_{\alpha\beta} f_{i\beta} f^\dagger_{j\gamma}
\sigma^\mu_{\gamma \rho} f_{j\rho} \cr\cr &\sim& - 2J_{ij}
\hat{\Delta}^\ast_{ji} \hat{\Delta}_{ji}   + \mathrm{Const} ,
\cr\cr \hat{\Delta}_{ji} &=&
\varepsilon_{\alpha\beta}f_{j\alpha}f_{i\beta}\,. \label{spinrep}
\eeqn

The QBT $d+id$ spin liquid state studied in this paper corresponds
to the following mean-field ansatz: \beqn \langle \hat{\Delta}_{j,
j+\hat{e}} \rangle \equiv \Delta_{j, j+\hat{e}} = \Delta^{(m)}
(e_x + i e_y)^2\, . \label{ansatz} \eeqn Here, $\Delta^{(m)}$
denotes first- ($m=1$) and second-neighbor ($m=2$) pairing
amplitudes, and $\hat{e}$ are the corresponding unit vectors on
the triangular lattice.

Expanded at $\vec{k} = 0$, the low-energy mean-field Hamiltonian
reads \beqn H_0 &\sim& \eta^t \left\{ - \sigma^y \lambda^x
(\partial_x^2 - \partial_y^2) + 2 \sigma^y \lambda^z \partial_x
\partial_y \right\} \eta\,. \label{MF} \eeqn This mean-field
Hamiltonian has quadratic band-touching (QBT) at $\vec{k} = 0$,
and the gauge symmetry is broken down to $Z_2$: $\eta \mapsto -
\eta$.

For a finite range around $\Delta^{(2)}/\Delta^{(1)}\simeq 0$, in
addition to the quadratic band touching at $\vec{k} = 0$, there
are also Dirac fermions at the Brillouin zone corners $\vec{Q} =
\pm (4\pi/3, 0)$. A complex Dirac fermion field $\chi$ at momentum
$\vec{Q} = (4\pi/3, 0)$ can be defined as \beqn \eta_{\vec{r}} =
\chi_{ \vec{r}} e^{i\vec{Q}\cdot \vec{r}} + \chi^\dagger_{\vec{r}}
e^{-i\vec{Q}\cdot \vec{r}}. \eeqn The low-energy Hamiltonian for
$\chi$ reads \beqn H_{\chi} \sim \chi^\dagger(- i \sigma^y
\lambda^x
\partial_x - i \sigma^y \lambda^z \partial_y )\chi\,.
\eeqn

The spinon carries a projective representation of the physical
symmetry group, and under discrete symmetry transformations, the
low-energy fields $\eta$ and $\chi$ transform as
\begin{eqnarray}
  T_x &:& x \mapsto x+1,
  \ \ \ \ \eta \mapsto \eta, \ \ \ \ \chi \mapsto e^{i4\pi/3}
  \chi; \cr\cr T &:& t \mapsto -t, \ \ \ \ \eta \mapsto i
  \lambda^y \eta, \ \ \ \ \chi \mapsto i\lambda^y \chi^\dagger;
  \cr\cr \mathcal{I} &:& \vec{r} \mapsto - \vec{r}, \ \ \ \ \eta
  \mapsto \eta, \ \ \ \ \chi \mapsto \chi^\dagger; \cr\cr
  \mathrm{P}_y &:& x \mapsto -x, \ \ \ \ \eta \mapsto
  i\sigma^y \lambda^x
  \eta, \ \ \ \ \chi \mapsto i\sigma^y\lambda^x \chi^\dagger; \cr\cr R_{\pi/3}
  &:& (x+iy) \mapsto e^{i\pi/3}(x+iy), \ \ \ \ \eta \mapsto
  e^{i\frac{\pi}{3} \lambda^y} \eta, \cr\cr && \chi \mapsto
  e^{i\frac{\pi}{3} \lambda^y}\chi^\dagger\,.
\end{eqnarray}
Importantly, notice that our QBT $d+id$ state is invariant under
all the discrete symmetries, including time reversal, reflection,
rotation, etc.

With these symmetries, no fermion bilinears are allowed to be
added to the low-energy mean-field Hamiltonian. For example, a
fermion bilinear $\eta^t \lambda^y \eta \sim f^\dagger f$ at the
$\Gamma$-point would gap out the QBT producing a chiral spin
liquid, but it is not allowed by time-reversal symmetry. On the
other hand, a marginally relevant four-fermion interaction can
lead to \emph{spontaneous} time-reversal symmetry breaking and
generate an exponentially small gap $m \eta^t \lambda^y \eta$ (see
also the main text).

Finally, we mention that on a finite sample, the QBT spin liquid
exhibits a pair of dispersionless edge modes at the mean-field
level. However, we expect that these edge modes will be unstable
when weak interactions are taken into account
\cite{Potter2013arXiv1303.6956P}.

\section{B: Details and discussion of our VMC calculations}

For the variational Monte Carlo (VMC) results presented in this
paper, we typically used $\sim$$10^6$ equilibrium sweeps, and
averaged over $2000$ spin configurations obtained from
$\sim$$10^6$ measurement sweeps. In estimating the error of our
measurements, we used a binning analysis with $\sim$$20$ samples
per bin.  In our plots, all errors are either on the order of or
smaller than the symbol size, or explicitly depicted with
one-sigma error bars.

To avoid degeneracies or singularities in the spin liquid wave
functions, we work with a mean-field Hamiltonian, Eq.~(2) of the
main text, with spinon boundary conditions periodic in one
direction and antiperiodic in the other direction; this ensures a
spin wave function with fully periodic boundary conditions. We
have carefully checked that our measurements, e.g.,
$\langle\vec{S}_i\cdot\vec{S}_j\rangle$ and $\langle
P_{ijkl}+\mathrm{H.c.}\rangle$, are converged in the system size
$L_x\times L_y$ and do not depend on the spinon boundary
conditions. The final energetics data presented in Fig.~2 of the
main text, as well as in Figs.~\ref{fig:exchanges_comp} and
\ref{fig:energy_cont} below, was taken on a $30\times30$ lattice.

\begin{figure}[b]
\begin{center}
\subfigure{\includegraphics[height=0.48\columnwidth]{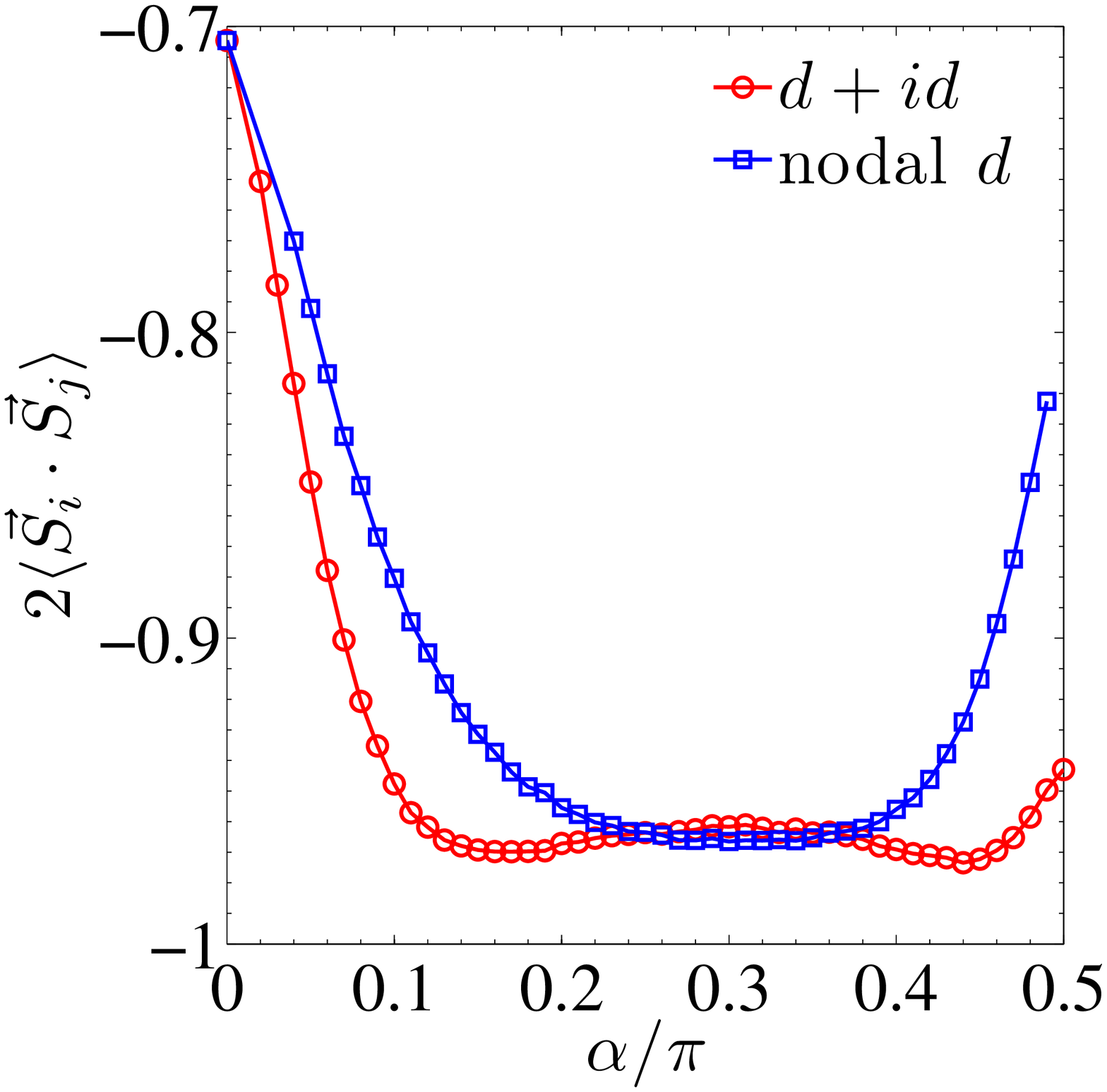}}
\hspace{0.0cm}
\subfigure{\includegraphics[height=0.48\columnwidth]{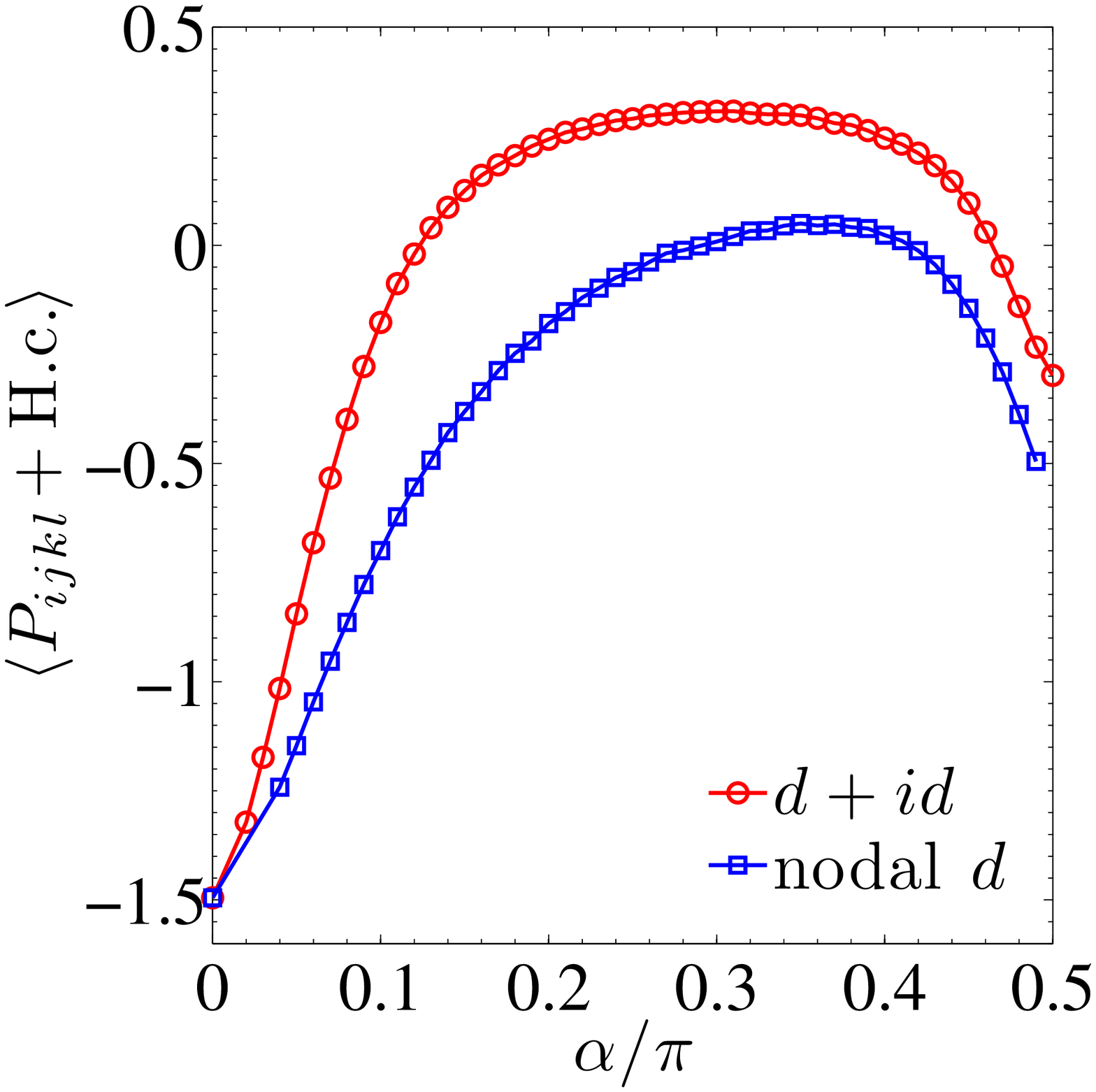}}
\end{center}
\vspace{-0.4cm} \caption{First-neighbor Heisenberg exchange (left)
and ring exchange (right) expectation values per site for the $d +
id$ and nodal $d$-wave states as a function of $\alpha =
\tan^{-1}(\Delta/t)$.  In the case of $d+id$, the special QBT
point lies at $\alpha/\pi=0.5$.} \label{fig:exchanges_comp}
\end{figure}

\begin{figure}[b]
\begin{center}
\subfigure{\includegraphics[height=0.49\columnwidth]{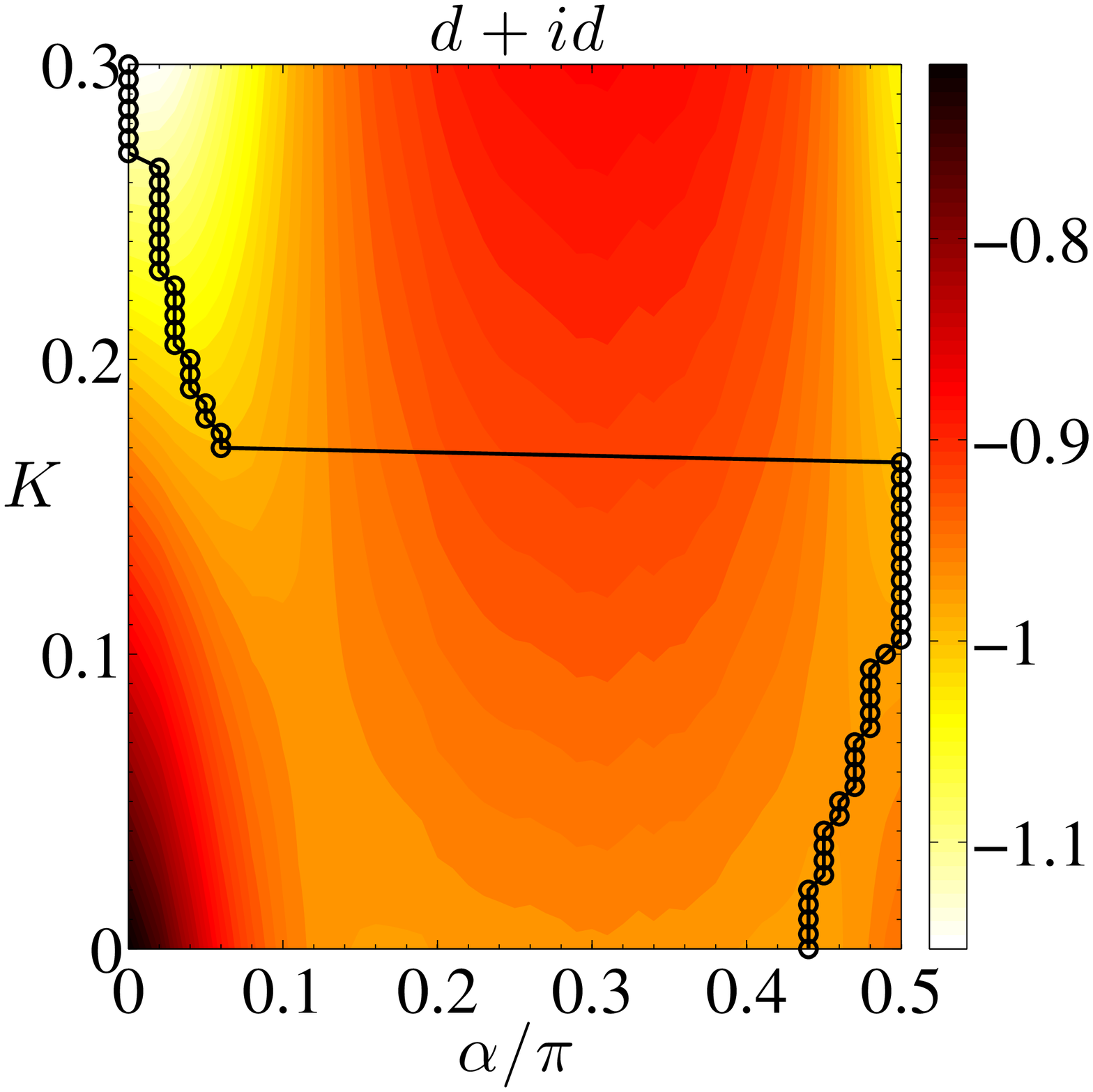}}
\hspace{-0.1cm}
\subfigure{\includegraphics[height=0.49\columnwidth]{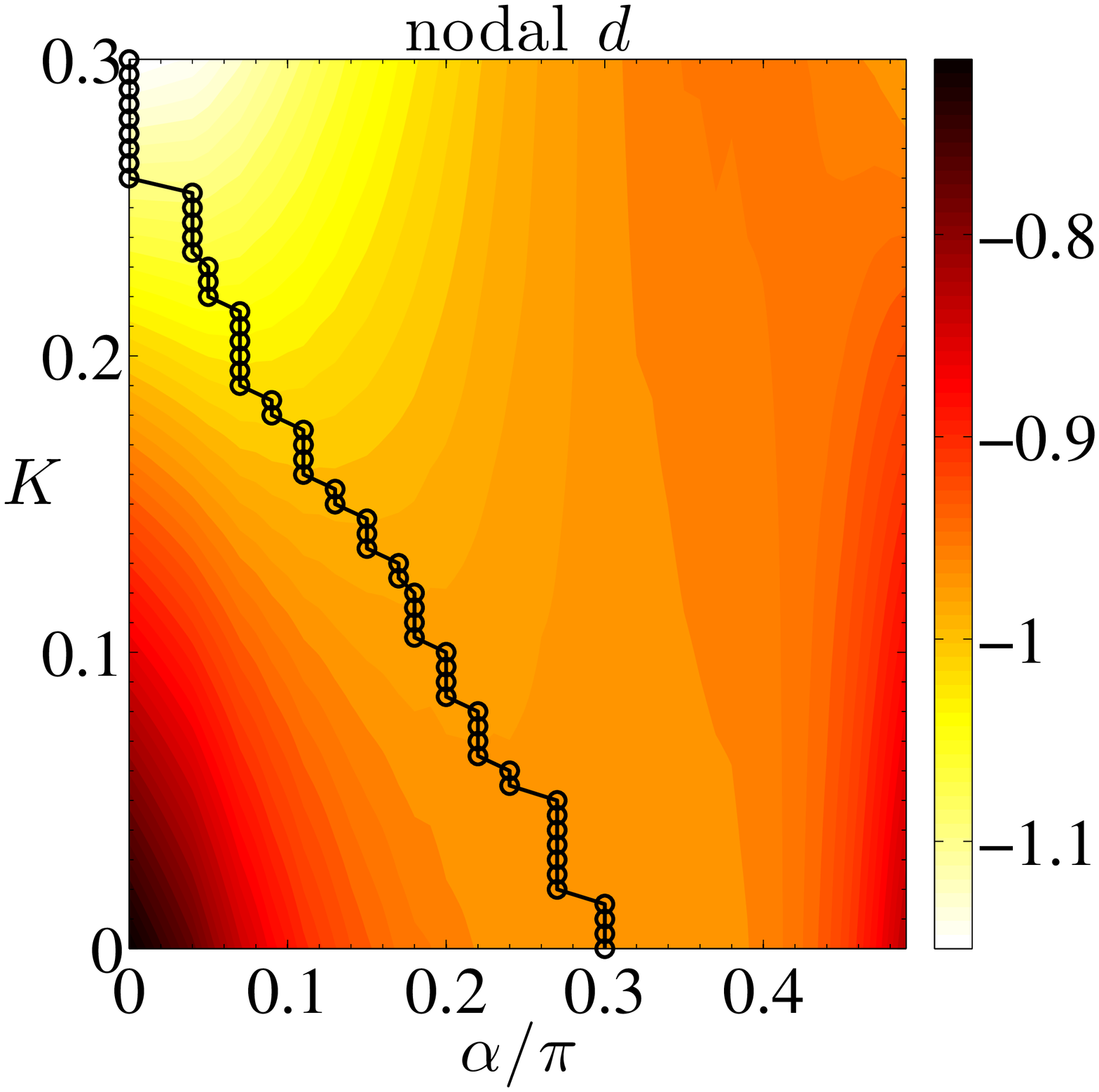}}
\end{center}
\vspace{-0.4cm} \caption{Contour plots of the trial energy per
site versus $K$ and $\alpha = \tan^{-1}(\Delta/t)$ for the $d+id$
(left) and nodal $d$-wave (right) states ($J_2=0$). The black
curves indicate the optimal $\alpha$ for each $K$ (see also Fig.~2
of the main text, bottom panel).} \label{fig:energy_cont}
\end{figure}

For the ordered states, we use three-sublattice $120^\circ$
antiferromagnetic (AFM) \cite{Huse88_PRL_60_2531,
Motrunich05_PRB_72_045105} and four-sublattice
\cite{Dagotto90_PRB_42_4800} states that are suggested within a
spin-wave analysis of the model at $K=0$.  Such ground state
ordering patterns (120$^\circ$ AFM and collinear) were also found
in exact diagonalization calculations on small lattice clusters
\cite{Dagotto90_PRB_42_4800,Lecheminant95_PRB_52_6647,Misguich99_PRB_60_1064,LiMing00_PRB_62_6372}.
Furthermore, we have checked the possibility of spiral orders with
arbitrary wave vectors (commensurate with the system size). We
have not considered the non-planar state proposed in
\cite{PhysRevLett.79.2081, Messio11_PRB_83_184401}, leaving this
for future work.

We note that quadratic fermion Hamiltonians similar to Eq.~(2) of
the main text can be used to construct states exhibiting magnetic
order \cite{GrosRVB, Bieri12_PRB_86_224409}. However, it turns out
that such Gutzwiller projected spin-density wave states do not
give good variational energies. In our case of triangular-lattice
spin-$1/2$ antiferromagnets, much better wave functions can be
obtained by applying spin-Jastrow factors to product states, as
pioneered in \cite{Huse88_PRL_60_2531}. We use such Huse-Elser
wave functions with first- and second-neighbor Jastrow factors to
compare their energies with those of the spin liquid states, and
to map out the variational phase diagram of the model
\cite{noteSU3}.

The variational phase diagram we obtain for the parameter range
and states considered is shown in Fig.~1 of the main text. On the
line $K=0$, we find that the $120^\circ$ AFM state is stable up to
$J_2\lesssim 0.05$, at which point a nodal $d$-wave spin liquid
starts to have the lowest energy within our set of wave functions.
It should be noted that on the line $K=0$, spin-wave calculations
predict a transition from the $120^\circ$~AFM state to a collinear
phase at $J_2= 0.125$ \cite{Dagotto90_PRB_42_4800}. Exact
diagonalization results seem to support a scenario without an
intermediate spin liquid phase \cite{Lecheminant95_PRB_52_6647}.
It is therefore possible that our variational approach
overestimates the spin liquid regions with respect to the ordered
states, especially in the case of the nodal $d$-wave spin liquid.
Also, we have not considered spin liquids obtained by spinon
hopping patterns with finite fluxes, as the ring term strongly
disfavors such states \cite{Motrunich05_PRB_72_045105} (although
Motrunich \cite{Motrunich04_unpublished} has found the
time-reversal invariant ``$U1B$'' state with alternating 0 and
$\pi$ fluxes through the triangles to be a promising ground state
candidate for $K\simeq0$ and $J_2\gtrsim0$). Interestingly, the
zero-flux U(1) spin liquid at $K\gtrsim 0.25$ is quite stable with
respect to second-neighbor interaction $J_2$, and it is not
destroyed within the parameter range and states we considered.

Finally, on the line $J_2=0$,
Motrunich~\cite{Motrunich05_PRB_72_045105,Motrunich04_unpublished}
found a transition to a spin liquid phase at $K\gtrsim 0.14$. This
small discrepancy with the transition point we find in the present
paper may be traced back to the fact that we did not keep as many
variational parameters in our Huse-Elser wave functions, and that
he considered a more restricted set of spin liquid states. We
leave a systematic study of the effect of more parameters in all
our wave functions for future work. However, we expect that the
spin liquid phase space may only slightly shrink, and that the
basic conclusions of our study remain unchanged.

Let us briefly discuss other variational quantum spin liquid wave
functions that we have also considered, but that were not
mentioned in the main text. We have generalized $s$- and $d$-wave
states by allowing three independent real singlet pairings
$\Delta_{ij}$ on first-neighbor links. Furthermore, we have
checked the possibility of finite-momentum pairing instabilities
of the U(1) spin liquid as proposed in \cite{Lee07_PRL_98_067006}
with the ``Amperean pairing'' state. However, we could not find
convincing evidence that such states are realized in the model,
Eq.~(1) of the main text.

Finally, in order to further improve the energy of the QBT $d+id$
state, we set the hopping $t_{ij}=0$ and added a second-neighbor
pairing amplitude $\Delta^{(2)}$ to the trial Hamiltonian, Eq.~(2)
of the main text. However, we found that the optimal parameter is
always essentially $\Delta^{(2)}/\Delta^{(1)} \simeq 0$ for the
model considered.

\section{C: Details of the energy landscape for the $d+id$ and nodal $d$-wave states}

We can gain further intuition for the energetics (at $J_2=0$) of
the first-neighbor $d+id$ and nodal $d$-wave states by considering
the $\alpha$-dependence [$\alpha=\tan^{-1}(\Delta/t)$] of
expectation values 2$\langle\vec{S}_i\cdot\vec{S}_j\rangle$ and
$\langle P_{ijkl}+\mathrm{H.c.}\rangle$ used to compute the
variational energy per site:
$E=J_1\left[2\langle\vec{S}_i\cdot\vec{S}_j\rangle\right]+K\left[\langle
P_{ijkl}+\mathrm{H.c.}\rangle\right]$. These results are shown in
Fig.~\ref{fig:exchanges_comp}.  We find the striking result that
the two local minima in the $d+id$ ansatz discussed in the main
text are actually present already in the pure Heisenberg model,
with the \emph{large}-$\Delta$ state ($\alpha/\pi\simeq0.44$,
$\Delta/t\simeq5.2$) slightly lower in energy (see
Fig.~\ref{fig:exchanges_comp}, left panel). Furthermore, we see in
the right panel of Fig.~\ref{fig:exchanges_comp} that beyond some
large value of $\Delta$ ($\alpha/\pi\simeq0.4$), we can gain
substantial ring energy in both the $d+id$ and nodal $d$-wave
states by actually further increasing $\Delta$. In the case of
$d+id$, going to the extreme limit $\Delta/t\rightarrow\infty$
loses only somewhat marginal Heisenberg energy while at the same
time gains significant ring energy:  It is ultimately a balance
between these two effects which makes the QBT $d+id$ state highly
competitive in the intermediate parameter regime of the ring
model.

Finally, in Fig.~\ref{fig:energy_cont} we show contour plots of
the trial energy versus both $K$ and $\alpha$ (for which the inset
of the bottom panel of Fig.~2 of the main text is a cross
section). For the $d+id$ state (left panel), we can clearly see
two basins of local minima: one of which connects to the QBT state
at $\alpha=\pi/2$, the other of which connects to the U(1) state
at $\alpha=0$. Upon increasing $K$ out of the pure Heisenberg
model, the large-$\Delta$ $d+id$ state quickly tracks to the QBT
state near $K\simeq0.1$, at which it remains until $K\simeq 0.17$
where the optimal $d+id$ state dramatically changes to one with
small $\Delta$. These results clearly show that the QBT $d+id$
state is a qualitatively new phase, and that it is not
continuously connected to the U(1) Fermi sea state at $\Delta=0$.
In sharp contrast, the optimal nodal $d$-wave state (right panel)
is continuously connected to the U(1) state for all values of $K$.

\bibliographystyle{prsty}
\bibliography{QBTdid_biblio}

\end{document}